\documentstyle{article}
\begin{document}
\begin{center}
{\Large \bf Exact analytical expression for magnetoresistance using quantum groups.}
\end{center}
\begin{center}

\textbf{S. A. Alavi  $^{\dag}$}\\

\textit{Department of Physics, Tarbiat Moallem university of Sabzevar, Sabzevar, P. O. Box 397,
 Iran}
\end{center}
\begin{center}
\textbf{S. Rouhani}\\
\textit{Department of Physics, Sharif University of Technology, Tehran,
P.O.Box 11365-9161, Iran.\\
Institute for studies in Theoretical Physics and Mathematics,
Tehran, 19395-5746, Iran.}

\end{center}


\emph{We obtain an exact analytical expression for
magnetoresistance using noncommutative geometry and quantum groups.Then we will show
that there is a deep relationship between magnetoresistance and
the quantum group $su_{q}(2)$, from which we understand the
quantum interpretation of
the quantum corrections to the conductivity.}\\

 \textbf{Introduction.} 
 Quantum groups [1] and deformed algebras have proved to be so rich and powerful that it
  seems natural to apply them to different problems in physics and mathematical physics. Many physicists 
  have studied the notion of quantum groups and deformed algebras from different points of view and 
  applied them to a variety of physical theories from nuclear and high energy physics to condensed 
  matter physics [See e.g. 2-15]. \\
 Quantum transport phenomena and magnetotransport in a two-dimensional
electron gas (2DEG) has been attracting much attention for scores
of years. This attention has been motivated by the progress in
preparing high-quality semiconductor heterostructures which has
opened up new areas in both fundemental and applied physics. The quantum corrections to the Drude
conductivity in disordered metals and doped semiconductors has
been most intensively studied for the last 20 years, see [16] for a review.\\ 
  The negative magnetoresistance induced by the suppression of the
quantum interference correction by magnetic field is a famous
manifestation of weak localization.\\
 
$\overline{\dag.Corresponding}$ $\overline{author.}$\\
E-mail:$Alavi@sttu.ac.ir$ and  
       $seyedalialavi@yahoo.com$\\ 

It gaves simple analytical expression for quantum correction to conductivity which allowed to
determine the phase breaking time experimentally.
A standard fitting procedure has been used to analyze the experimental data
and determine the phase breaking time [17,18].
The phase breaking time is the fitting parameter.Another approach to study the
negative magnetoresistance due to weak localization has been
presented in Ref[19], which is based on a quasi-classical treatment
of the problem [20,21,22], and an analysis of the statistics of
closed paths of a particle moving over 2D plane with randomly
distributed scatterers. This method has been used to study the
weak localization in InGaAs/GaAs heterostructures with single[23]
and double [24] quantum wells.\\
As mentioned above there is an analytical expression which gives the
dependence of the conductivity correction on the magnetic field [17].\\
\begin{equation}
\Delta\sigma(B)=\delta\sigma(B)-\delta\sigma(0)=aG_{0}[\Psi(0.5+\frac{B_{tr}}{B}\gamma)-\ln(\frac{B_{tr}}{B}\gamma)],
\end{equation}
where $G_{0}=\frac{e^{2}}{2\pi^{2}\hbar}$,$B_{tr}=\frac{\hbar
c}{2e\ell^{2}}$,$\gamma=\frac{\tau}{\tau_{\varphi}}=\frac{\ell}{\ell_{\varphi}}$
,$\ell$ is the mean free path,$\ell_{\varphi}$ is the phase
breaking length,$\Psi(x)$ is a diagamma function,$\tau$
and $\tau_{\varphi}$ are elastic and phase breaking time
respectively. The value of prefactor $"a"$ is theorectically equal
to unity. Using this expression physicists have been attempting to
analyse the experimental data and to determine phase breaking
time or phase breaking length and their temperature dependence
through the fitting of the experimental curves. It is shown in
Ref [23] that  Eq.(1) describes the magnetic field dependence of
the magnetoresistance relatively well but with
prefactor $a<1$.\\
In order to give an expression for conductivity correction in a
magnetic field Minkov et al. [19], introduced the distribution of
closed random paths of area $"s"$ $w_{N}(s)$ such that
$w_{N}(s)ds$ gives the probability density of return after $N$
collisions following a trajectory which enclosed the area in the
range $(s,s+ds)$, this gives:
\begin{equation}
\delta\sigma(b)=-2\pi G_{0}\{\ell^{2}\sum_{N=3}^{\infty}
\int_{-\infty}^{+\infty} ds e^{-\frac{L}{\ell_{\varphi}}} w_{N}(s)
\cos\frac{(1+\gamma)^{2}bs}{\ell^{2}}\},
\end{equation}
where $b=\frac{B}{(1+\gamma)^{2}B_{tr}}$. They took into acount
inelastic processes destroying the phase coherence by including
the factor $e^{-\frac{L}{\ell_{\varphi}}}$, where L is the path
length. Note that the expression in the bracket is dimensionless,
and it doesn't make any difference what units of measurement we choose
for length and area. Therefore we can treat
$\ell,\ell_{\varphi},L$ and $s$ as dimensionless variables. We
introduce new dimensionless
variables $\alpha=\ell^{2}$ and $\beta=\ell\ell_{\varphi}$ which we will use them later.\\
In this paper which is in close relation with [12], we provide an exact expression for $w_{N}(s)$ using
noncommutative geometry and quantum groups. Then by use of  it we present an exact
analytical expression for magnetoresistance. we will also discuss
the relationship between magnetoresistance and the
 $su_{q}(2)$ algebra.\\

\textbf{2.Exact area distributions of closed random paths.}\\
 Bellissard et al.[25] obtained the exact area distributions of
 closed random walks using noncommutative geometry. Their method
 was based upon Harper model [26]. Let us consider a spinless
 electron on a two dimensional lattice and submitted to a uniform
  magnetic field along the z-direction and perpendicular to the
 plane of motion. The system is not invariant under translations
 but there is an invariance under the so-called magnetic
 translation operators $w(a)$. The Harper model Hamiltonian is
 [26]:
\begin{equation}
H=\sum_{\mid a\mid =1}w(\vec{a})=w(\vec{a})+w(\vec{b})+w(-\vec{a})+w(-\vec{b}),
\end{equation}
Here $\vec{a}$ and $\vec{b}$ are two perpendicular unit vectors which build the lattice  unit cell, $\mid a\mid=\mid b\mid=1$.
 After some calculations they showed that the probability distribution of closed random paths is given by [25]:
\begin{equation}
P(A,N)=(\frac{2\pi
N}{4^{N+1}})\frac{1}{\sqrt{2\pi}}\int_{-\infty}^{+\infty} e^{-iax}
T(H^{N}(x)) dx ,
\end{equation}
  where $a=\frac{A}{N}$ is the renormalized area, $A$ is the algebraic area
 and $N$ is the number of collisions. $\eta=\frac{x}{N}$ is a dimensionless variable and $\eta=2\pi \frac{\Phi}{\Phi_{0}}$. $\Phi$ and $\Phi_{0}$ are the magnetic flux through the unit cell and the quantum of flux respectively.  $H$ is the Harper's model
 Hamiltonian and its trace is given by [25]:
\begin{equation}
T(H^{N}(x))=\frac{4^{N+1}}{2\pi
N}\frac{\frac{x}{4}}{\sinh(\frac{x}{4})}[1-\frac{1}{2N}\frac{(\frac{x}{4})^{2}}{\sinh^{2}(\frac{x}{4})}+O(\frac{1}{N^{2}})]
,
\end{equation}

  In the limit of large $N$, when the diffusion approximation is
 valid, we can neglect the second term, and equ.(4), (5) and the 
 normalization condition (which ensures that the total probability 
  is equal to unity) give:
\begin{equation}
P_{N}(A)=\frac{1}{N}\frac{\pi}{\cosh^{2}(\frac{2\pi A}{N})} .
\end{equation}

 To make connection with equ.(2), note that:
\begin{equation}
w_{N}(A)=\frac{1}{\pi N}P_{N}(A) .
\end{equation}

 Compare with the equ.(13) and (15) in [19].\\

 \textbf{3. Exact analytical expression for magnetoresistance.}\\
 Now we can write equ.(2) as follows:
\begin{equation}
\delta\sigma(b)=-2\pi G_{0}\{\alpha \sum_{N=3}^{\infty}
\int_{-\infty}^{+\infty} dA e^{-\frac{N}{2\beta}} w_{N}(A)
\cos(\frac{(1+\gamma)^{2}bA}{\alpha})\} .
\end{equation}
 Using equs.(4),(5),(7)and (8)we will obtain the following
 expression for magnetoresistance:\\

 $\Delta\sigma(B)=\delta\sigma(B)-\delta\sigma(0)=$ \\
\begin{equation}
-\sqrt{\frac{2}{\pi}}\alpha G_{0}\sum_{N=3}^{\infty}
e^{-\frac{N}{2\beta}}\{[1-\frac{1}{2N}\frac{(\frac{BN}{4\alpha
B_{tr} })^{2}}{\sinh^{2}(\frac{BN}{4\alpha B_{tr}
})}]\frac{(\frac{B}{4\alpha B_{tr} })}{\sinh(\frac{BN}{4\alpha
B_{tr} })}-\frac{1}{N}(1-\frac{1}{2N})\}.
\end{equation}

 Our appropriate candidate for exact distribution of closed random walks using quantum groups presented in [12] 
  leads to the following exact expression for magnetoresistance:\\

$\Delta\sigma(B)=\delta\sigma(B)-\delta\sigma(0)=$ \\
\begin{equation}
-\sqrt{\frac{2}{\pi}}\alpha G_{0}\sum_{N=3}^{\infty}
e^{-\frac{N}{2\beta}}\{\cos^{N}(\frac{(\frac{B}{4\alpha B_{tr} })}{\sinh(\frac{BN}{4\alpha
B_{tr} })})\frac{(\frac{B}{4\alpha B_{tr} })}{\sinh(\frac{BN}{4\alpha
B_{tr} })}-\frac{1}{N}\cos^{N}(\frac{1}{N})\}.
\end{equation}
   
  which strongly supports the equ.(9). For comparison we have plotted $\Delta\sigma(B)/G_{0}$ as given by equ.(1), (9)
 and  equ.(10) for a given values of $\ell$ and $\ell_{\varphi}$ in fig.1. We have taken into
 account $a=1$. In the summation of equ.(9) and equ.(10) we have allowed $N$ to
 range from $3$ to $900$. As it is seen the difference between equ.(9) and equ.(10) is negligibly small and equ.(9) is also exact. 
 From physical point of view they are the same but from mathematical point of view there is a small 
 difference between them, note that :
\begin{equation}
\cos^{N}(\frac{x}{N})=1-\frac{1}{2N}x^{2}+O(x^{4}).
\end{equation}

It is worth to mention that there are two other theoretical results [18,22] for quantum corrections to the
 conductivity . In the diffusion approximation i.e.  when the number of collisions is much greater than
  unity, we have [18] :\\
  
  \hspace{.6cm}$\Delta\sigma(b)=\delta\sigma(b)-\delta\sigma(0)=
  aG_{0}[\Psi(0.5+\frac{\gamma}{b})-\Psi(0.5+\frac{1}{b})-\ln(\gamma)]$\\

   For $x>>1$, $\Psi(0.5+x)\simeq ln(x)$, and we get equ.(1). 
  The calculatios of $\Delta\sigma(B)$ beyond the diffusion approximation show that $\Delta\sigma(B)$ 
  deviates from this equation if the number of collisions for actual trajectories is not very large [18].
   The role of nonbackscattering contribution to magnetoresistance has been studied in Ref.[22]. They 
   showed that the enhancement of backscattering responsible for the weak localization is accompanied by 
	 a reduction of the scattering in other directions. The reduction of the scattering at the arbitrary angles
 	  leads to the decrease of the quantum correction to the conductivity. Within the diffusion approximation 
	  this decrease is small, but it should be taken into account in the case of a relatively strong magnetic 
 	   field. They have performed numerical calculations whics shows that the inclusion of nonbackscattering 
		 contribution leads to a decrease in magnetoconductance.\\

  \textbf{4. Comparison with experimental data.}\\
 We compared equ.(10)(or equ.(9)) with the experimental data of structure $1$
 samples in fig.2. The heterostructures with $200 A^{0} In_{0.07}Ga_{0.93}$, as quantum
 well, $\delta$-doped by $Si$ in the centre. It should be mentioned that we have considered
 $\beta$ as fitting parameter and we have $\ell_{\varphi}=\frac{\beta}{\sqrt{\alpha}}$. As it is seen,
 exact expression is in good agreement with experiment and on the
 other hand in our approach there is no need to introduce the
 imposed prefactor $"a"$, and the number of fitting parameters is
 reduced. The temperature dependence of $\ell_{\varphi}$ is plotted in fig.3. As
 is observed the exact analytical expression suggests $\ell_{\varphi}\propto T^{p}$ with
 $p=-1$.\\
 As mentioned in Ref[27], at low temperatures the phase-breaking
 time is determined by inelasticity of the electron-electron
 interaction and is:
\begin{equation}
\tau_{\varphi}=\frac{\hbar}{kT}\frac{\sigma_{0}}{2\pi
G_{0}}\frac{1}{\ln(\frac{\sigma_{0}}{2\pi G_{0}})},
\end{equation}
where $\sigma_{0}=\frac{e^{2}k_{F}\ell}{2\pi\hbar}$, $G_{0}=\frac{e^{2}}{2\pi^{2}\hbar}$, and therefore: $\frac{\sigma_{0}}{\pi G_{0}}=k_{F}\ell$     , then we have:\\
\begin{equation}
\tau_{\varphi}=\{\frac{k_{F}\ell}{\ln(\frac{k_{F}\ell}{2})}\}\frac{\hbar}{2kT}
.
\end{equation}
 The expression in the bracket is not important, because it is a number
and can be omitted by changing the scale of measurement. The
results of the fitting of $\tau_{\varphi}$ obtained from exact and
approximate(i.e. equ.1) expressions with $\frac{c}{T}$ (fig.3), are
as follows:
\begin{equation}
c_{exact}=2.3764352
\end{equation}
\begin{equation}
c_{approx}=2.2059803
\end{equation}
The one obtained from equ.(12) is:
\begin{equation}
c=2.3687788
\end{equation}
As it is seen the exact expression is in very good agreement with
the equ.(12).\\

\textbf{5. Connection between magnetoresistance and the quantum
groups.}\\
Quantum algebras are the q-deformation of the ordinary Lie
algebras [1]. Our argument is based on the quantum algebra
$su_{q}(2)$. The generators of the $su_{q}(2)$ algebra satisfy the
commutation relations:
\begin{equation}
[j_{3},j_{\pm}]=\pm j_{\pm} .
\end{equation}
\begin{equation}
[j_{+},j_{-}]=\frac{q^{2j_{3}}-q^{-2j_{3}}}{q-q^{-1}} .
\end{equation}.\\
One can show that the following combinations of the magnetic
translations:
\begin{equation}
j_{+}=\frac{w(\vec{a})+w(\vec{b})}{q-q^{-1}} ,
\end{equation}
\begin{equation}
j_{-}=-\frac{w(-\vec{a})+w(-\vec{b})}{q-q^{-1}} ,
\end{equation}
where  $w(\vec{b}-\vec{a})=q^{j_{3}}$  satisfy the
$su_{q}(2)$ algebra. $q$ is the parameter of deformation, $q=\exp(i\frac{e}{\hbar}\vec{B}.(\vec{a}\times\vec{b}))=e^{2\pi i\frac{\Phi}{\Phi_{0}}}$.
  The magnetic translation operators $w(\vec{a})$ satisfy the following relation:\\
\begin{equation}
w(\vec{a}) w(\vec{b})=\exp(i\frac{e}{2\hbar}\vec{B}.(\vec{a}\times\vec{b}))w(\vec{a}+\vec{b}) .
\end{equation}
 On the other hand from equs.(3), (19) and (20) we have:
\begin{equation}
H=(q-q^{-1})(j_{+}-j_{-})=2i(q-q^{-1})j_{y} .
\end{equation}
Hence  Harper's Hamiltonian is a generator of  the $su_{q}(2)$
algebra. From equs.(4),(7) and (8) we have:
\begin{equation}
\delta\sigma(B)=-\sqrt{\frac{2}{\pi}}G_{0}\alpha\sum_{N=1}^{\infty}
\frac{1}{N} e^{-\frac{N}{N_{0}}}T(H^{N}(x))\mid_{x=\lambda N} ,
\end{equation}
where $N_{o}=2\beta$. We have changed the lower limit in the
summation from $3$ to $1$, because the paths with $N=1,2$ have
zero areas and therefore have zero contributions to the
conductivity corrections. We can omit $\frac{1}{N}$ because it
will be eliminated if we calculate $T(H^{N}(x))$ at $x=\lambda N$
, where $\lambda =\frac{B}{\alpha B_{tr}}$. Therefore we have:
\begin{equation}
\delta\sigma(B)=-\sqrt{\frac{2}{\pi}}G_{0}\alpha
T\{\sum_{N=1}^{\infty}[(q-q^{-1})
e^{-\frac{1}{N_{0}}}(j_{+}-j_{-})]^{N}\} ,
\end{equation}
or\\
\begin{equation}
\delta\sigma(B)=-\sqrt{\frac{2}{\pi}}G_{0}\alpha
T\{\frac{1}{1-(q-q^{-1}) e^{-\frac{1}{N_{0}}}(j_{+}-j_{-})}\} .
\end{equation}
 Expression in the bracket is a Green function, and this is a
key point to understand the quantum interpretation of the quantum
correction to the conductivity. This can be written as:
\begin{equation}
\delta\sigma(B)=-\sqrt{\frac{2}{\pi}}G_{0}\rho\alpha
T\{\frac{1}{(j_{-}+\frac{\rho}{2})-(j_{+}-\frac{\rho}{2})}\} ,
\end{equation}

with: $\rho =e^{\frac{1}{N_{0}}}(q-q^{-1})^{-1}$. This equation shows the relationship between conductivity 
 corrections in a magnetic field and the $su_{q}(2)$ algebra and reflects the quantum symmetry 
  of magnetoresistance.\\
In conclusion the quantum group is a key symmetry of the magnetoresistance problem and has as its 
 consequence the exact analytical expression for magnetoresistance. It if found here that(see also [12]), the 
  Harper Hamiltonian is a generator of the $su_{q}(2)$ algebra and there is a closed connection between 
   random walks problem in two dimensions and the $su_{q}(2)$ algebra. \\

\textbf{ Acknowledgments.}\\
We would like to thank Professor K. Esfarjani and 
 Professor R. Ghorbanzadeh for their  careful reading of the manuscript and for their valuable comments.
  We acknowledge M. Ebrahimi for helpful discussions.\\
  
  \textbf{ Figure Captions}\\
Fig1. Magnetic field dependence of $\Delta\sigma(B)/G_{0}$ as given by equ.(1)(dashed lines)
, equ.(10)(solid line) and equ.(9)(dots). $B_{tr}=0.52 T$ and the values of dimentionless variables $\alpha=\ell^{2}$
 and the fitting parameter $\beta=\ell\ell_{\phi}$(i.e. $\ell_{\varphi}=\frac{\beta}{\sqrt{\alpha}}$) are $\alpha=1.33$
  and $\beta=12.7$.\\
Fig2. Magnetic field dependence of $\Delta\sigma(B)/G_{0}$ for temperatures $T=1.5 K,2 K,2.5$\\
$K,3 K$
, and $4.2 K$. 
Solid lines are the results of fitting by exact expression(equ.(10) 
 or equ.(9)). 
Circles are experimental data for structures 1. Dashed line shows the result 
 of fitting by equ.(1). $B_{tr}=0.52 T$, $\alpha=1.33$ and $\beta=26.71, 22.5, 19.2, 16.5, 12.7$ for the temperatures $T= 1.5 K, 2. K,
  2.5 K, 3. K$ and $4.2 K$ respectively.\\
 Fig3.Temperature dependence of $\tau_{\phi}$. Solid lines are the dependencies $\tau_{\phi}=\frac{C}{T}$
  with $C_{exact}$ and $C_{approx}$ respectively. Circles and plus are the results of fitting 
  by equ.(10)(or equ.(9)) and equ.(1) respectively. Open squares are obtained from equ.(12).\\
 \textbf{References. }\\
1. M. Cahichian, A. Demichev, Introduction to Quantum groups, World Scientific, Singapore, 1996.
 V. G. Drinfeld in Proc. ICM Berkeley, CA, ed
A. M. Gleason(Providence, RI: AMS) P 798 1986.; M. Jimbo
Lett. Math. Phys 1063, 1985; Lett. Math. Phys 11 247, 1986. P. Kulish
and Yu. Reshtikhin, J. Sov. Math. 23(1983) 2435.\\
2. M. Cahichian, P. Kulish, Phys. Lett. B 234 (1990) 72.\\
3. H.-T. Sato, Mod. Phys. Lett. A 9 (5) (1994) 451, (20) (1994) 1819.\\
4. A. M. Gavrilik, hep-ph/9712411.\\
5. D. Bonatsos, C. Daskaloyannis, Prog. Part. Nucl. Phys. 43 (1999) 537.\\
6. S. Shelly Sharma, N. K. Sharma, nuc-th/0009048.\\
7. R. J. Finkelstein, Mod. Phys. Lett. A 15 (2000)1709.\\
8. S. Alam, Phys. Lett. A 272 (2000) 107.\\
9. J. K. Slingerland, F. A. Bais, Nucl. Phys. B 612 (2001) 229.\\
10. M. Chaichian, P. Presnajder, hep-th/0209024.\\
11. A. Jellal. Mod. Phys. Lett. A 17 (2002) 701.\\
12. S. A. Alavi, M. Sarbishaei, Phys. Lett. A. 299 (2002) 116.\\
13. M. Tierz, hep-th/0308121.\\
14. P. Pouliot, hep-th/ 0306261.\\
15. M. Arik, S. Gun, A. Yldz, Eur. Phys. J. C 27 (2003) 453.\\      
16. P. A. Lee and T. V. Ramakrishnan, Rev. Mod. Phys., 57, 287(1985),
G. Bergman, Physics Peports, 107, 1(1984), B.L.Altshuler et al.,
 in Quantum theory of solids, edited by I.M.Lifshitz,(Mir Publication, Moscow,1982). 
 B.L.Altshuler and A.G.Aronov, in Electron-Electron Interaction
in Disordered systems, edited by A.L.Efros and M.Pollak,(North
Holland, Amsterdam,1985) P.1. B. D. Simons and A. Altland, in Theoretical Physics at the End of the XXth Century, Banff, CRM Series in 
   Mathematical Physics (Springer, New York, 2001), Chap. Mesoscopic Physics.\\
 17. S.Hikami,A.Larkin and Y.Nagaoka, Prog.Theor.Phys. 63,707(1980).\\
18. H.-P. Wittman and A.Schmid, J.Low.Temp.Phys. 69, 131(1987).\\
19. G.M.Minkov et al., Phys. Rev. B. 61, 13164(2000).\\
20. S. Chakravarty and A. Schmid, Phys. Reports 140, 193(1986).\\
21. M. I. Dyakonov, Solid St.Comm. 92, 711(1994)\\
22. A. P. Dmitriev et al., Phys. Rev. B 56, 9910 (1997).\\
23. G. M. Minkov et al., Phys.Rev.B 61, 13172(2000).\\
24. G. M. Minkov et al., Phys.Rev.B. 62, 17089(2000).\\
25. J. Bellissard et al., J.Phys.A: Math. Gen. 30 (1997) L 707-709.\\
26. P. G. Harper, Proc. Phys. Soc. Lon. A 68 874 and 879.\\
27. G. M. Minkov et al., Cond-mat/0104299.\\
 
\end{document}